\documentclass[twocolumn,superscriptaddress, showpacs] {revtex4}
\usepackage{graphicx}
\usepackage{amsmath}
\usepackage{subfigure}
\usepackage{dsfont}
\usepackage{amssymb}
\usepackage{float}
\begin{document}

\title[Short Title]{Dissipative preparation of tripartite singlet state in coupled arrays of cavities via quantum feedback control}
\author{X. Q. Shao\footnote{E-mail: shaoxq644@nenu.edu.cn}}
\affiliation{Center for Quantum Sciences and School of Physics, Northeast Normal University, Changchun, 130024, People's Republic of China}
\affiliation{Center for Advanced Optoelectronic Functional Materials Research, and Key Laboratory for UV Light-Emitting Materials and Technology
of Ministry of Education, Northeast Normal University, Changchun 130024, China}
\author{Z. H. Wang}
\affiliation{Center for Quantum Sciences and School of Physics, Northeast Normal University, Changchun, 130024, People's Republic of China}
\affiliation{Center for Advanced Optoelectronic Functional Materials Research, and Key Laboratory for UV Light-Emitting Materials and Technology
of Ministry of Education, Northeast Normal University, Changchun 130024, China}
\author{H. D. Liu}
\affiliation{Center for Quantum Sciences and School of Physics, Northeast Normal University, Changchun, 130024, People's Republic of China}
\affiliation{Center for Advanced Optoelectronic Functional Materials Research, and Key Laboratory for UV Light-Emitting Materials and Technology
of Ministry of Education, Northeast Normal University, Changchun 130024, China}
\author{X. X. Yi}
\affiliation{Center for Quantum Sciences and School of Physics, Northeast Normal University, Changchun, 130024, People's Republic of China}
\affiliation{Center for Advanced Optoelectronic Functional Materials Research, and Key Laboratory for UV Light-Emitting Materials and Technology
of Ministry of Education, Northeast Normal University, Changchun 130024, China}

\begin{abstract}
We propose an experimentally feasible scheme for dissipative preparation of tripartite entangled state with atoms separately trapped in an array of
three coupled cavities. The combination of coherent driving fields and quantum-jump-based feedback control will drive the system into a non-equilibrium steady state, which has a nearly perfect overlap with the genuine three-atom singlet state. Different control strategies are investigated and the corresponding optimal parameters are confirmed. Moreover, the fidelity of target state is insensitive to detection inefficiencies, and it oversteps 90\% for a wide range of decoherence parameters as long as the single-atom cooperativity
parameter $C\equiv g^2/(\gamma\kappa)>350$.
\end{abstract}
 \maketitle \maketitle
\section{Introduction}
The concept of quantum entanglement lies at the heart of quantum information sciences, whose feature was so peculiar as to be termed spooky action at a distance by Einstein \cite{ein}. The most simple examples of entanglement are the Bell states \cite{Bell}, which construct a complete basis for bipartite quantum states and are maximally entangled measured by concurrence or other computable quantities \cite{woo,vidal}. For tripartite or multipartite quantum systems, the maximally entangled states generally relate to two non-equivalent classes of
entangled states, i.e. Greenberger-Horne-Zeilinger (GHZ) states and $W$ states \cite{ghz,wstate}. Compared with bipartite entanglement, the correlations in multipartite entanglement are more subtle, which can be utilized in quantum cryptography, communication complexity, and other quantum information tasks. For instance, the cluster states have been shown to
constitute a universal resource for one way quantum
computation proceeding only by local measurements
and feedforward of their outcomes \cite{HJ}.

Cabello for practical use reasons introduced the $N$-particle $N$-level singlet states $|S_N\rangle$ \cite{Cabe}.
These states are $N$ particles of spin-$(N-1)$/2 with total spin
zero, which are $N$-lateral rotationally invariant and can provide correlated result for measurement of spin. Due to these properties, they were considered as efficient
solutions to $N$-strangers, secret
sharing and liar detection problems. In addition, the decoherence-free subspaces robust
against collective decay can be constructed from the supersinglet states \cite{Cabe1}, which further widens the scope of these quantum states in quantum information processing. It should be noted that there has no experimental report on generating  supersinglet states for $N\geq 3$ yet, while some theoretical proposals are exclusively based on the unitary dynamics in closed systems \cite{Jin,Lin,shao,chenz}.
In fact, a practical quantum system cannot be isolated from the environment: on the one hand, there is an inevitable interaction between quantum system and its surroundings, which is the origin of decoherence; on the other hand, the intrinsic property of a quantum system will not be revealed unless a series of measurements are performed by external apparatus. Therefore, the theory of open quantum system becomes a research hot spot, since it provides a more realistic image for characterizing evolution of quantum states.

The primary idea about dissipation-induced entanglement was put forward in Ref.~\cite{ple}, it was then further exploited to generate entanglement of distinct optical cavities and atoms from white noise \cite{ple1,yi}. Although
the amount of entanglement was so small, these works changed people perception of dissipation, i.e. the environment can be a kind of resource to prepare entanglement. Recently, Kastoryano {\it et al}. proposed a theoretical scheme for preparation of a bipartite entangled state in a leaking optical cavity, and a nearly pure singlet state can be obtained with unity probability from arbitrary initial state \cite{kastoryano}. This method improves the amount of entanglement significantly and paves the way for generation of various bipartite entanglement both in theory and in experiment \cite{busch,shen,torre,rao,carr,lin,zz,shaoxq,ben,gon}. Nevertheless, as the number of particles increases, it is hard to derive an effective master equation for open quantum system due to the complexity of dynamic evolution. Fortunately, quantum feedback control will provide an alternative way to manipulate the quantum system during the evolution \cite{1,2}, e.g. the quantum-jump-based feedback control has successfully worked on preparation of four-qubit decoherence-free subspace \cite{3}.

In this paper, we propose an experimentally feasible scheme for dissipative preparation of tripartite singlet state $|S_3\rangle$ with quantum feedback control. Three atoms are separately trapped in an array of three-coupled cavities, which is convenient to address qubit individually. The quantum feedback operations are applied right after the photons leaking out of cavities detected by three detectors, respectively. The coherent driving fields combined with quantum-jump-based feedback control will finally drive the system into a nearly pure stationary state, irrespective of initial state. The prominent advantage of our scheme is that it relaxes the bad-cavity limit, and the fidelity of the target  surpasses 90\% for a wide range of decoherence parameters on condition that the single-atom cooperativity
parameter $C=g^2/\gamma\kappa>350$.
\section{effective hamiltonian}
\begin{figure}
\scalebox{0.28}{\includegraphics{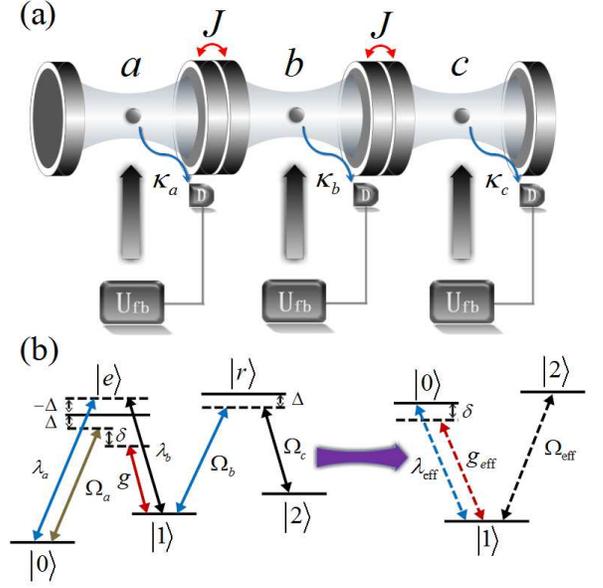} }
\caption{\label{p1}(Color online) Schematic view of the setup and the
configuration of atoms. The system consists of three $M$-type atoms trapped in three coupled cavities, respectively.
Each atom is driven by five dispersive laser fields and simultaneously coupled to its own cavity mode, which is then reduced to an effective $V$-type atom. The photon can hop between neighboring cavities with coupling
strength $J$. Different control strategies are applied to the atoms on
the basis of output for the leaky cavities.}
\end{figure}

The system consists of three atoms with $M$-type configuration
trapped in a coupled-cavity array, as shown in Fig.~\ref{p1}.
Each atom interacts with its own cavity mode via the Jaynes-Cummings model with the coupling constants $g$ and detuning $\Delta$. Additionally, there are five classical fields driving the transition between levels $|e\rangle(|r\rangle)\leftrightarrow|0\rangle (|2\rangle)$ and $|1\rangle$, with Rabi frequencies $\lambda_a^m$, $\lambda_b^m$, $\Omega_a^m$, $\Omega_b^m$, $\Omega_c^m$, and detuning $\mp\Delta$, respectively.
The photon can hop between neighboring  cavities with coupling strength $J$. All the parameters are assumed to be real in the context.
In the interaction picture, the Hamiltonian of the system reads ($\hbar = 1$)
\begin{eqnarray}\label{a1111}
{\cal H}_{\rm I}&=&\sum_{i=1}^3\bigg[\Omega_a^{i}|e\rangle_{ii}\langle 0|e^{i\Delta t}+
\Omega_b^{i}|r\rangle_{ii}\langle 1|e^{i\Delta t}\nonumber\\
&&+\Omega_c^{i}|r\rangle_{ii}\langle 2|e^{i\Delta t}+\lambda_a^i|e\rangle_{ii}\langle 0|e^{-i\Delta t}\nonumber\\
&&+\lambda_b^i|e\rangle_{ii}\langle 1|e^{-i\Delta t}\bigg]
+g(a|e\rangle_{11}\langle 1| +b|e\rangle_{22}\langle 1|
\nonumber\\&&+c|e\rangle_{33}\langle 1|)e^{i(\Delta+\delta) t}+J(a^{\dag}b+b^{\dag}c)+{\rm H.c.}.
\end{eqnarray}
In order to investigate dynamics of the system further, we introduce three normal bosonic modes $c_1=(a-c)/{\sqrt{2}}$,
$c_2=(a-\sqrt{2}b+c)/2$, and
$c_3=(a+\sqrt{2}b+c)/2$ corresponding frequencies $0$, $-\sqrt{2}J$, and $\sqrt{2}J$. These modes are not
coupled with each other, but interact with the atoms because
of the contributions of the cavity fields. Meanwhile, for a large detuning condition $|\Delta|\gg \{g,\Omega^i_{a,(b,c)},\lambda^i_{a,(b)},\delta\}$,
we may safely eliminate the excited states $|e\rangle$ and $|r\rangle$, then the Hamiltonian reduces to
\begin{eqnarray}\label{heff}
{\cal H}_{\rm eff}&=&\frac{g_{\rm eff}^{1}}{2}|0\rangle_{11}\langle 1|e^{i\delta t}
(c_3e^{-i\sqrt{2}Jt}+c_2e^{i\sqrt{2}Jt}+\sqrt{2}c_1)\nonumber\\
&&+\frac{g_{\rm eff}^{2}}{\sqrt{2}}|0\rangle_{22}\langle 1|e^{i\delta t}
(c_3e^{-i\sqrt{2}Jt}-c_2e^{i\sqrt{2}Jt})\nonumber\\
&&+\frac{g_{\rm eff}^{3}}{2}|0\rangle_{33}\langle 1|e^{i\delta t}
(c_3e^{-i\sqrt{2}Jt}+c_2e^{i\sqrt{2}Jt}-\sqrt{2}c_1)\nonumber\\
&&+\sum_{i=1}^3\big[\Omega_{\rm eff}^i|2\rangle_{ii}\langle 1|+\lambda_{\rm eff}^i|1\rangle_{ii}\langle 0|\big]+{\rm H.c.},
\end{eqnarray}
where $g^i_{\rm eff}=g\Omega_a^i/\Delta$, $\Omega^i_{\rm eff}=\Omega_b^i\Omega_c^i/\Delta$, and $\lambda^i_{\rm eff}=-\lambda_a^i\lambda_b^i/\Delta$.
The stark-shift terms of ground states are not considered here since they can be absorbed via introducing ancillary levels. Especially for our system, only one additional level is needed to cancel the energy shift of $|2\rangle$, because the stark-shift terms of other two states have been removed automatically through the arrangement of classical fields. The first three lines of Eq.~(\ref{heff}) describes three effective $V$-type atoms interacting with three cavity modes $c_1$, $c_2$, and $c_3$, with detuning $\delta$, $\delta+\sqrt{2}J$, and $\delta-\sqrt{2}J$, respectively. If the condition $\delta=\sqrt{2}J$ and $\delta\gg g^i_{\rm eff}$ are satisfied, we can implement a Hamiltonian that governs a selectively resonant interaction between atoms and mode $c_3$, i.e.
\begin{equation}\label{xxx}
{\cal H}_{\rm sr}=\sum_{i=1}^3\frac{G}{2}|2\rangle_{ii}\langle 0|c_3+\Omega|2\rangle_{ii}\langle 1|+\Omega|1\rangle_{ii}\langle 0|+{\rm H.c.},
\end{equation}
where we have supposed $g^1_{\rm eff}=g^3_{\rm eff}=\sqrt{2}g^2_{\rm eff}=G$, and $\Omega_{\rm eff}^i=-\lambda_{\rm eff}^i=\Omega$ for the sake of convenience. Eq.~(\ref{xxx}) makes our scheme much feasible with different experimental setups, since the coupling strengths are all adjustable by modulating detunings and Rabi frequencies of classical fields.

\section{dissipative dynamics}
The dissipative dynamics of current system is described by the Lindblad master equation
\begin{eqnarray}\label{444}
\dot{\rho}&=&-i\Omega[(J_1^{+}+J_1^-),\rho]-i\Omega[(J_2^{+}+J_2^-),\rho]-i \frac{G}{2}[(J_1^+c_3\nonumber\\&&+J_1^-c_3^{\dag}),\rho]+\frac{\kappa}{2}(2c_3\rho c_3^{\dag}-c_3^{\dag}c_3\rho-\rho c_3^{\dag}c_3)\nonumber\\
&=&{\cal L }\rho-i \frac{G}{2}[(J_1^{+}c_3+J_1^-c_3^{\dag}),\rho]+\kappa{\cal D}[c_3]\rho,
\end{eqnarray}
where $\kappa$ denotes the decay rate of cavity mode $c_3$, and the collective amplitude
damping operator has been defined as $J_1^-=\Sigma_{i=1}^3|1\rangle_{ii}\langle 0|$ and
$J_2^-=\Sigma_{i=1}^3|1\rangle_{ii}\langle 2|$. It should be clarified that we have substituted the effective Hamiltonian into master equation instead of the full Hamiltonian Eq.~(\ref{a1111}), partly because the excited states and other cavity modes are largely detuned to the system (the effect of spontaneous emission will be discussed in Sec.V), and partly because it is instrumental for us to clearly understand the dissipative dynamics of considered systems.

The dominate factor in Eq~(\ref{444}) is the decay rate of cavity $\kappa$, for the relation $\kappa\gg G$ is always attainable via
modulating the detuning between atom and cavity, even for a superstrong coupling regime of cavity quantum electrodynamics (QED).
Thus the highly excited modes may be
neglected in the limit of large decay rates, and the density matrix $\rho$ can be expanded in small photon number states to a good approximation \cite{wang}, i.e.
\begin{eqnarray}\label{222}
\rho&=&\rho_{0,0}|0\rangle_{c}\langle 0|+\rho_{1,0}|1\rangle_{c}\langle 0|+\rho_{0,1}|0\rangle_{c}\langle 1|+\rho_{1,1}|1\rangle_{c}\langle 1|\nonumber\\
&&+\rho_{2,0}|2\rangle_{c}\langle 0|+\rho_{0,2}|0\rangle_{c}\langle 2|+{\cal O}(G^3/\kappa^3),
\end{eqnarray}
where ${\cal O}$ represents the high-order small quantities, and $|n\rangle_c$ denotes the  state of cavity mode having $n$ photons. After substituting the above equation into Eq.~(\ref{444}) and neglecting terms of greater than second order,  we obtain
a set of coupled equations for the field-matrix
elements:
\begin{equation}\label{e1}
\dot{\rho}_{0,0}={\cal L}\rho_{0,0}-i \frac{G}{2}[J_1^+\rho_{1,0}-\rho_{0,1}J_1^-]
+\kappa\rho_{1,1},
\end{equation}
\begin{eqnarray}\label{e2}
\dot{\rho}_{1,0}&=&{\cal L}\rho_{1,0}-i \frac{G}{2}[J_1^-\rho_{0,0}-\rho_{1,1}J_1^-+\sqrt{2}J_1^+\rho_{2,0}]
\nonumber\\&&-\frac{\kappa}{2}\rho_{1,0},
\end{eqnarray}
\begin{equation}\label{e3}
\dot{\rho}_{1,1}={\cal L}\rho_{1,1}-i \frac{G}{2}[J_1^-\rho_{0,1}-{\rho}_{1,0}J_1^+]-\kappa\rho_{1,1},
\end{equation}
\begin{equation}\label{e4}
\dot{\rho}_{2,0}={\cal L}\rho_{2,0}-i \frac{G}{2}[\sqrt{2}J_1^-\rho_{1,0}]-\kappa\rho_{2,0}.
\end{equation}
Under the condition $\kappa\gg G$, it is reasonable to assume $\dot{\rho}_{1,0}=0$ and $\dot{\rho}_{2,0}=0$, and then get the values of of these operators as
\begin{equation}\label{8}
\rho_{1,0}=\rho_{0,1}^{\dag}\approx-\frac{iG}{\kappa}[J_1^-\rho_{0,0}-\rho_{1,1}J_1^-],
\end{equation}
\begin{equation}\label{8}
\rho_{2,0}=\rho_{0,2}^{\dag}\approx-\frac{iG}{\sqrt{2}\kappa}J_1^-\rho_{1,0}.
\end{equation}
By
substituting the corresponding results into Eqs.~(\ref{e1}) and (\ref{e3}), we obtain a diagonal density matrix whose elements are
\begin{eqnarray}\label{8}
\dot{\rho}_{0,0}&=&{\cal L}\rho_{0,0}-\frac{G^2}{2\kappa}[J_1^+J_1^-\rho_{0,0}+\rho_{0,0}J_1^+J_1^-\nonumber\\&&-2J_1^+\rho_{1,1}J_1^-]+\kappa\rho_{1,1},
\end{eqnarray}
\begin{eqnarray}\label{8}
\dot{\rho}_{1,1}&=&{\cal L}\rho_{1,1}-\frac{G^2}{2\kappa}[J_1^-J_1^+\rho_{1,1}+\rho_{1,1}J_1^-J_1^+\nonumber\\&&-2J_1^-\rho_{0,0}J_1^+]-\kappa\rho_{1,1}.
\end{eqnarray}
Adding these two equations together and adiabatically eliminating
the elements $\rho_{1,1}$, the effective
master equation for three atoms becomes
\begin{eqnarray}\label{999}
\dot{\rho}&=&{\cal L}\rho-\frac{G^2}{2\kappa}[J_1^+J_1^-\rho+\rho J_1^+J_1^--2J_1^-\rho J_1^+]\nonumber\\
&=&{\cal L}\rho+\Gamma{\cal D}[J_1^-]\rho,
\end{eqnarray}
where $\Gamma=G^2/\kappa$ is the collective amplitude
damping rate of the transition $|0\rangle\rightarrow|1\rangle$. A simple inspection shows that the tripartite singlet state $
|S_3\rangle=(|012\rangle
-|102\rangle-|210\rangle+|120\rangle+|201\rangle-|021\rangle)/{\sqrt{6}}
$ is a stationary state solution
of Eq.~(\ref{999}), and this is the reason why the supersinglet states can be a resource for constructing
 decoherence-free subspaces with respect to collective decoherence. Nevertheless, the dissipation along is not enough to drive an arbitrary initial state towards the tripartite singlet state, as $|S_3\rangle$ is not the unique steady
solution. In the following, we will particularly show the power of quantum feedback control on generation of quantum entanglement.
\section{quantum-jump-based feedback}
\begin{figure}
\scalebox{0.45}{\includegraphics{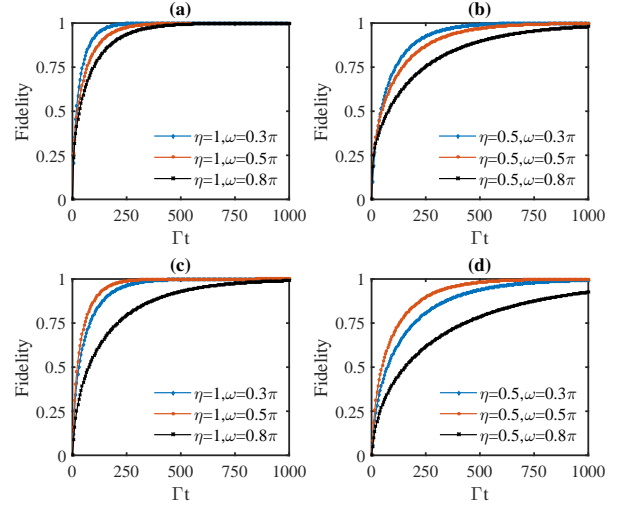} }
\caption{\label{p3}(Color online) The effects of detection efficiencies and feedback control parameters on the rate of convergence for target state, where the driving frequency $\Omega=0.5\Gamma$. For the nonlocal control, a choice of $\omega=0.3\pi$ guarantees a more than 99\% fidelity at a short time $\Gamma t=200$, while for the random local control, $\omega=0.5\pi$ is the optimal value. The detection inefficiencies will not decrease the fidelity of final state,
but delays the time at which stationarity is achieved.}
\end{figure}

Quantum feedback control is a class of methods to manipulate the system towards certain desired state taking advantage of system's quantum state or trajectory. Generally speaking,  a feedback signal is typically filtered or processed in a classical way, such as measurement based feedback, but keeps the quantum coherence of the output at the same time. In our scheme, each cavity output of the coupled arrays of cavities is monitored by a
photodetector $D$ whose signal provides the input to the application
of the control Hamiltonian. Thus the stochastic master equation can
be modified as
\begin{figure}
\begin{minipage}[t]{0.5\linewidth}
\centering
\includegraphics[width=1.8in]{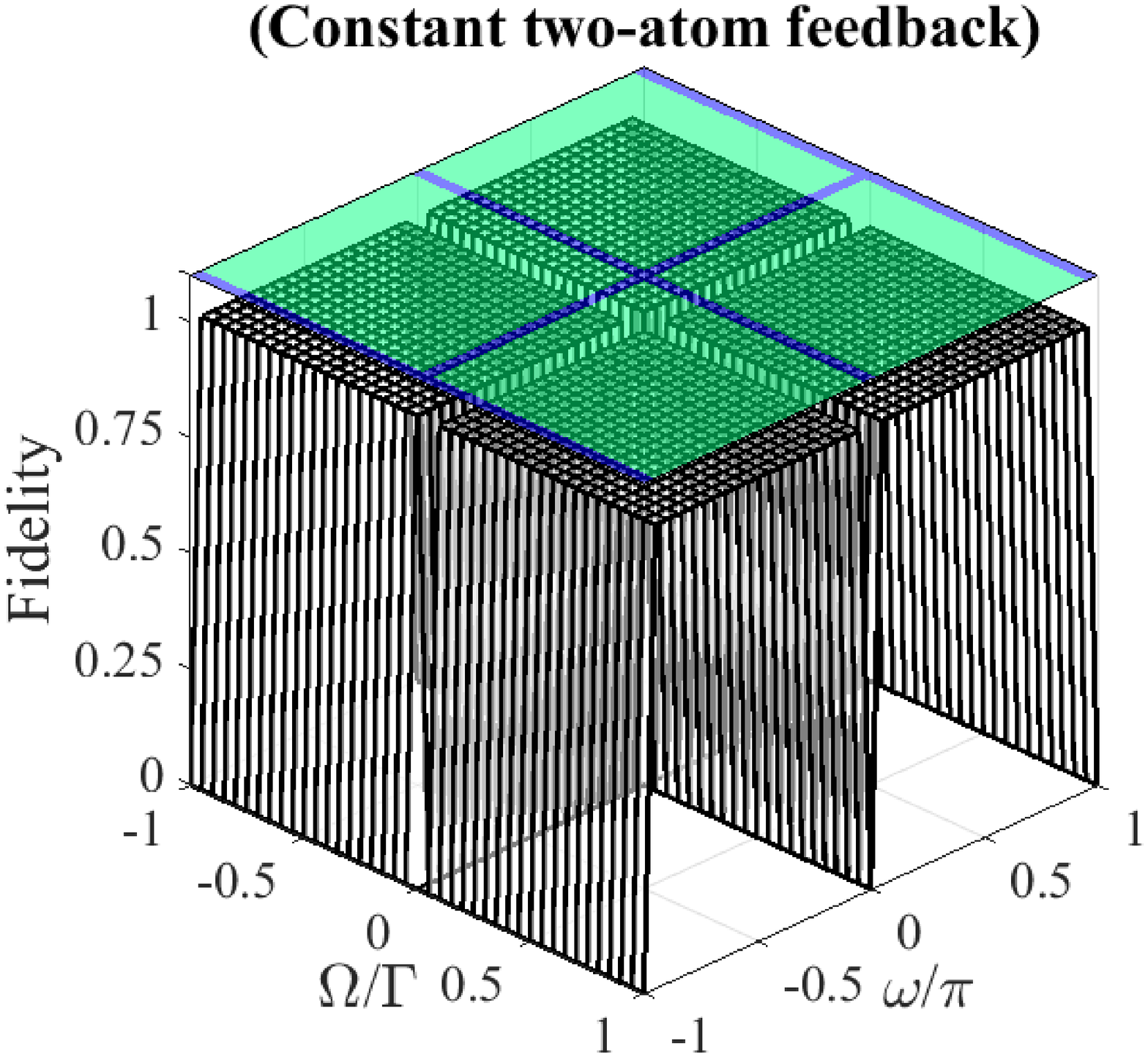}
\label{fig:side:a}
\end{minipage}%
\begin{minipage}[t]{0.5\linewidth}
\centering
\includegraphics[width=1.8in]{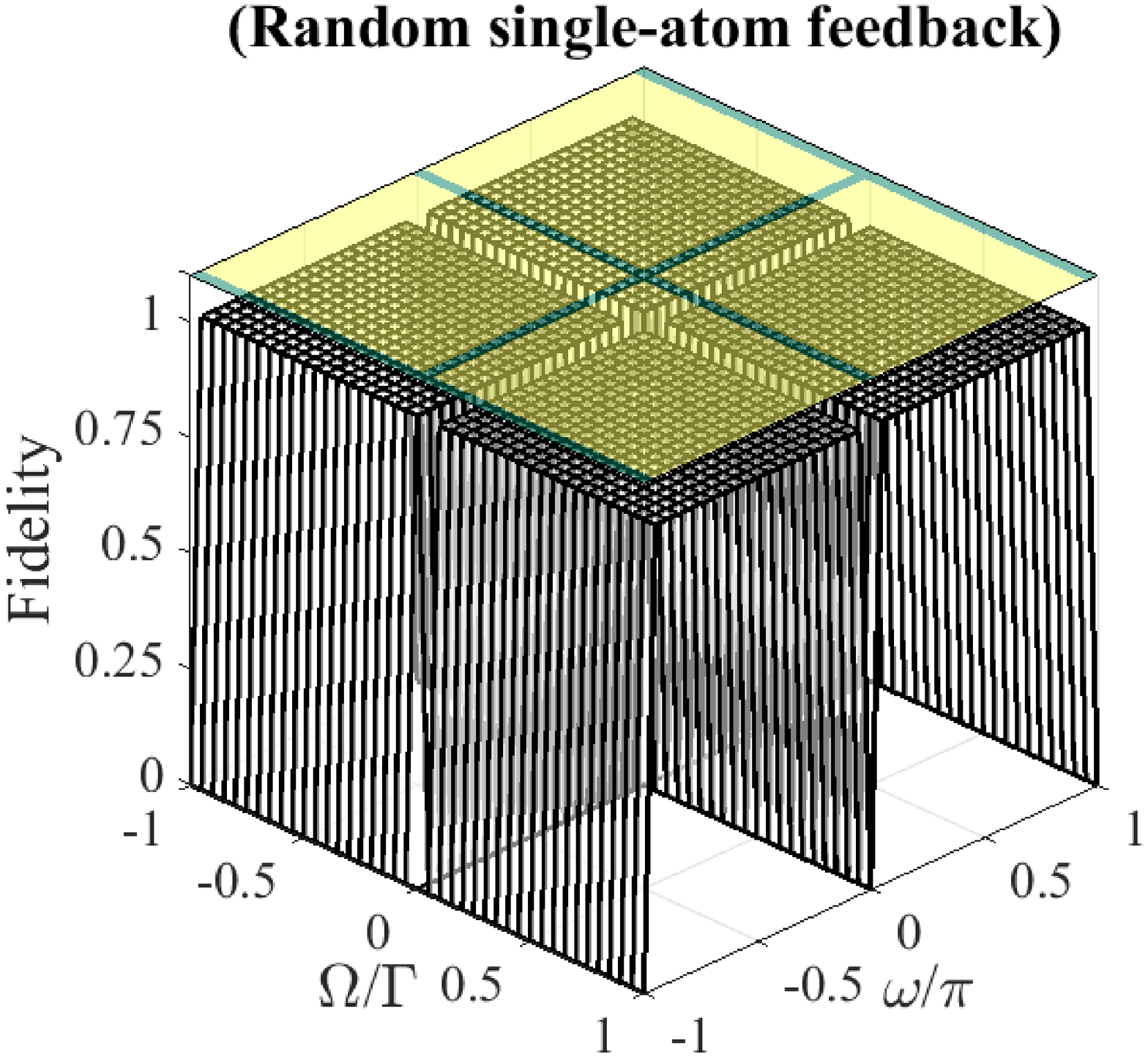}
\label{fig:side:b}
\end{minipage}
\begin{minipage}[t]{0.5\linewidth}
\centering
\includegraphics[width=1.8in]{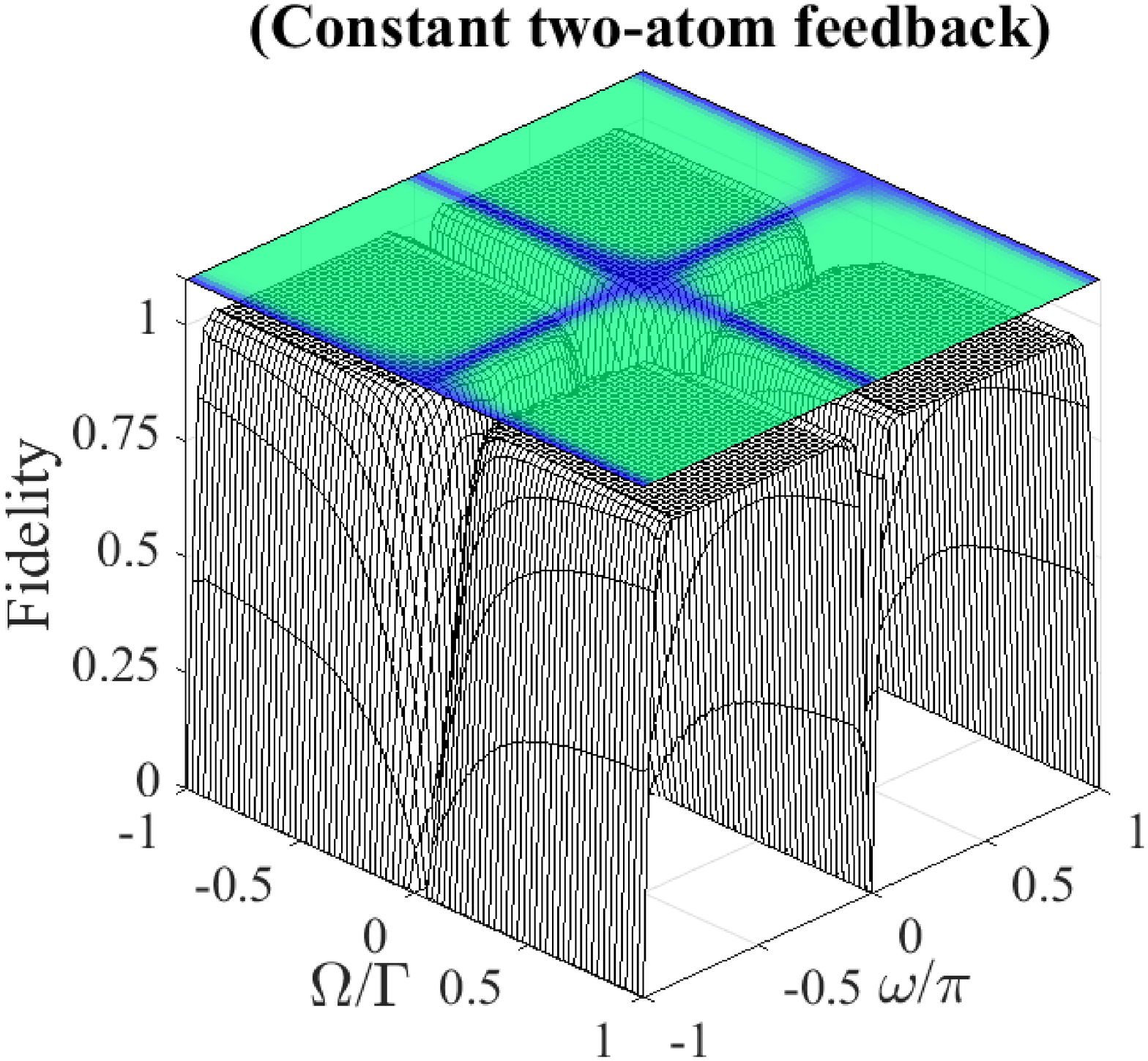}
\label{fig:side:a}
\end{minipage}%
\begin{minipage}[t]{0.5\linewidth}
\centering
\includegraphics[width=1.8in]{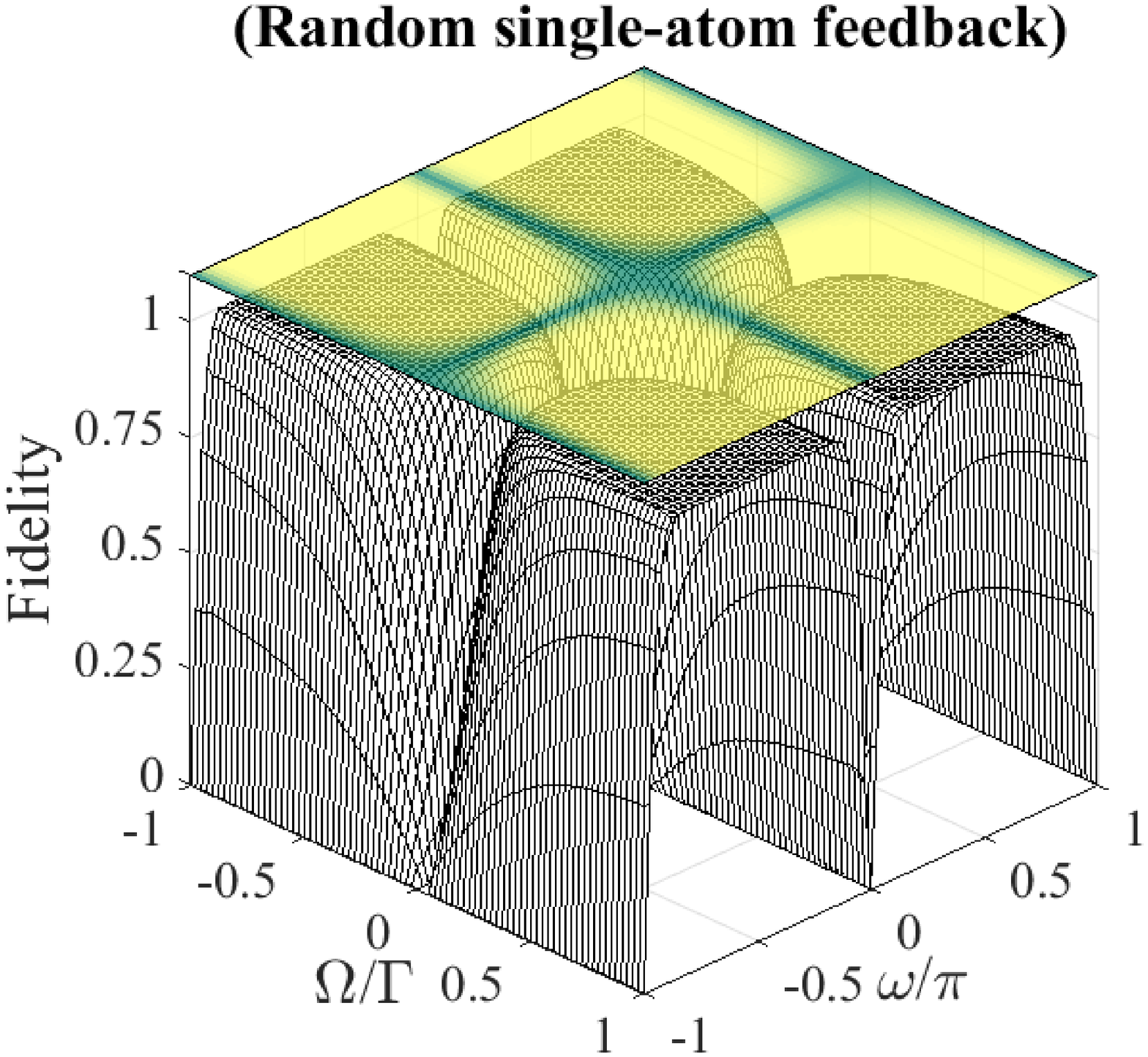}
\label{fig:side:b}
\end{minipage}
\caption{\label{p2}(Color online) The fidelities of steady state (top) and the final state after a long time evolution from a random initial state (bottom) versus the nonlocal (left) and local (right)
feedback control parameters and driving fields strength.}
\end{figure}
\begin{equation}\label{H10}
\dot{\rho}={\cal L }\rho\
+\Gamma{\cal D}[U_{{\rm fb}}J_1^-]\rho,
\end{equation}
where the jump feedback operator is unitary that maintains the stationary solution of $|S_3\rangle$ because of ${\cal D}[U_{{\rm fb}}J_1^-]=
U_{{\rm fb}}J_1^-\rho J_1^+U_{{\rm fb}}^{\dag}-(J_1^+J_1^-\rho+\rho J_1^+J_1^-)/2$. As the principle of selecting feedback control is to violate the symmetry with respect to exchange of atoms, we first choose $U_{{\rm fb}}=\exp[-i\omega(|1\rangle_{11}\langle 0|+|0\rangle_{11}\langle 1|)-2i\omega(|1\rangle_{22}\langle 0|+|0\rangle_{22}\langle 1|)]$, i.e. the quantum feedback is performed on two appointed atoms simultaneously right after a detection click, so we refer to this kind of control as a constant two-atom feedback, or a non-local feedback. In contrast, we may also implement the  feedback on single atom randomly  instead of the constant two-atom feedback, which corresponds to a local feedback. The unconditioned master equation for
this case is derived as
\begin{eqnarray}\label{H11}
\dot{\rho}&=&{\cal L }\rho
+\frac{1}{3}\Gamma{\cal D}[U_{{\rm fb}}^1J_1^-]\rho+\frac{1}{3}\Gamma{\cal D}[U_{{\rm fb}}^2J_1^-]\rho\nonumber\\&&+\frac{1}{3}\Gamma{\cal D}[U_{{\rm fb}}^3J_1^-]\rho,
\end{eqnarray}
where $U^i_{{\rm fb}}=\exp[-i\omega(|1\rangle_{ii}\langle 0|)]$$(i=1,2,3)$, and we have supposed the probabilities that the feedback acting on each atom are equal for simplicity. On the top of Fig.~\ref{p2}, we employ the definition of fidelity $F(|S_3\rangle,\rho_\infty)=\sqrt{\langle S_3|\rho_\infty|S_3\rangle}$ for steady state to assess the performance of different controlling strategies \cite{Nielsen}. These results illustrates that a unity fidelity is always achievable as long as the driving strength and feedback satisfying $\Omega*\omega/\Gamma\neq 0$. For realistic situation, it is meaningful to discuss the asymptotic fidelity $F(|S_3\rangle,\rho_t)$ at a finite time. Thus we also plot the fidelity of the final state after a long time evolution $\Gamma t=1500$ from an initial state $|111\rangle_{1,2,3}$ at the bottom of Fig.~\ref{p2}. The selections of optimal parameters are different in both cases, which can be seen more clearly in Fig.~\ref{p3}(a) and Fig.~\ref{p3}(c). For a fixed Rabi frequency of driving field $\Omega=0.5\Gamma$, a choice of $\omega=0.3\pi$ guarantees the fidelity exceeding 99\%  at a short time $\Gamma t=200$ in the non-local feedback control scheme, while for the random local control scheme, $\omega=0.5\pi$ is the optimal value. If the detection efficiency is not perfect, the master equations of Eqs.~(\ref{H10}) and (\ref{H11}) should be reformulated as
\begin{equation}\label{HHH}
\dot{\rho}={\cal L }\rho\
+\eta\Gamma{\cal D}[U_{{\rm fb}}J_1^-]\rho
+(1-\eta)\Gamma{\cal D}[J_1^-]\rho,
\end{equation}
and
\begin{eqnarray}\label{HHH}
\dot{\rho}&=&{\cal L }\rho
+\frac{1}{3}\eta\Gamma{\cal D}[U_{{\rm fb}}^1J_1^-]\rho+\frac{1}{3}\eta\Gamma{\cal D}[U_{{\rm fb}}^2J_1^-]\rho\nonumber\\&&+\frac{1}{3}\eta\Gamma{\cal D}[U_{{\rm fb}}^3J_1^-]\rho
+(1-\eta)\Gamma{\cal D}[J_1^-]\rho,
\end{eqnarray}
where $\eta$ represents the efficiency of the detector and $(1-\eta)$ corresponds to the case when the
detector fails to click and no feedback control is performed. Fig.~\ref{p3}(b) and Fig.~\ref{p3}(d)
characterize the evolution of fidelities for both constant two-atom feedback and random single-atom feedback with $\eta=0.5$. Compared with case of perfect detection,  the time of convergence for entanglement is delayed, but the fidelity of target state is not affected.
\section{effect of spontaneous emission}
\begin{figure*}
\scalebox{0.36}{\includegraphics{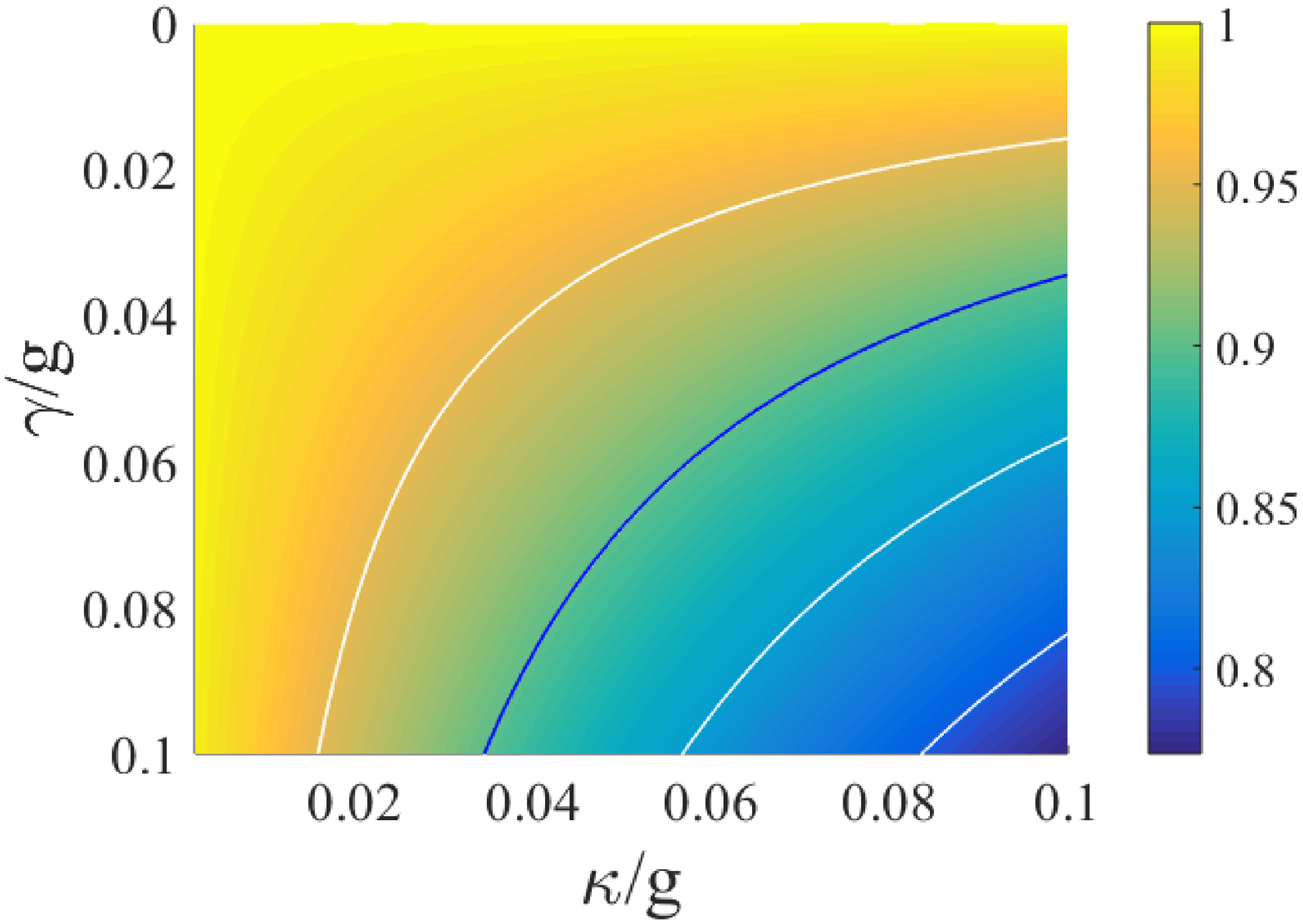} }
\scalebox{0.36}{\includegraphics{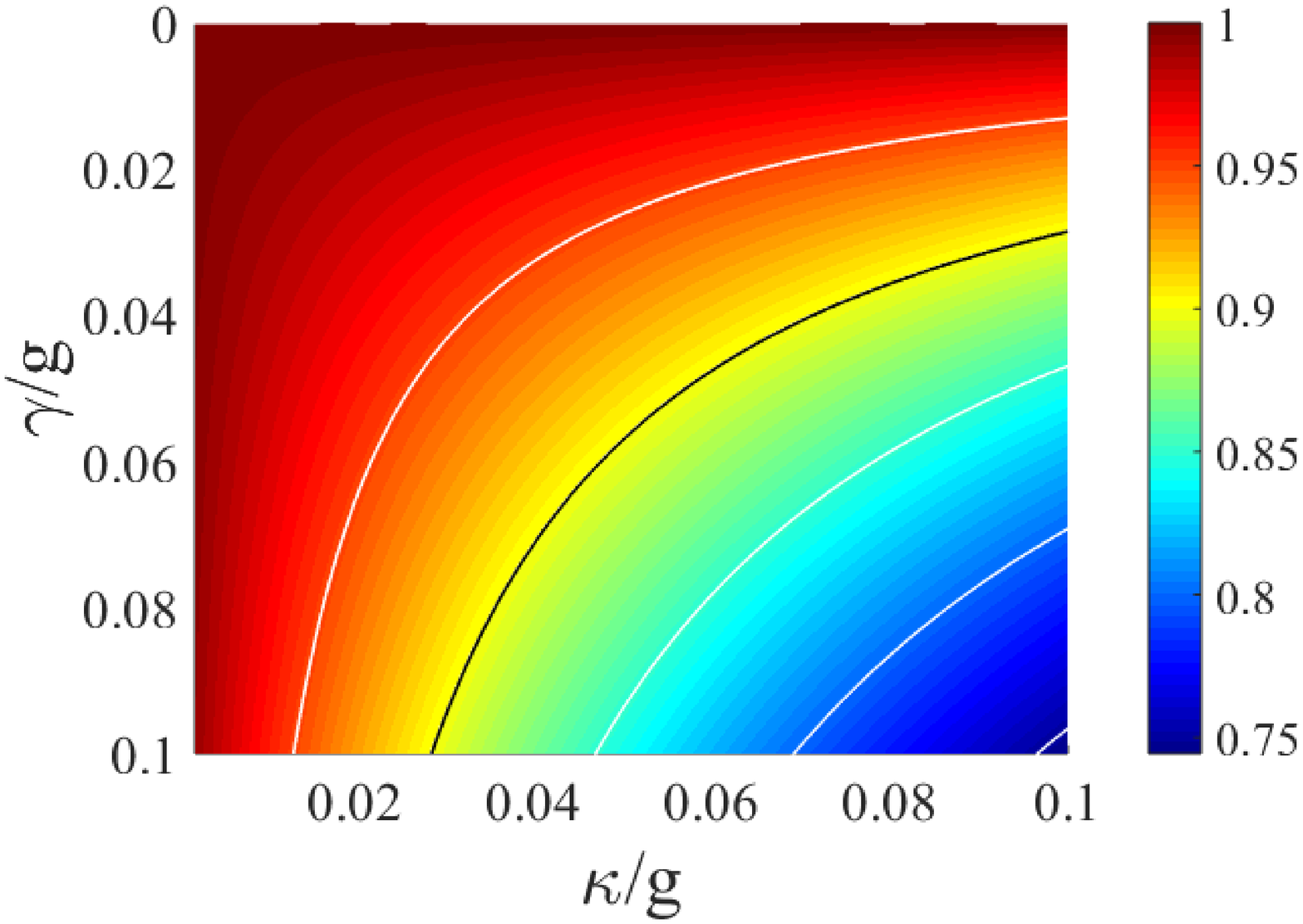} }
\caption{\label{p4}(Color online) The contour plots of fidelity for steady state versus spontaneous emission $\gamma/g$ and cavity decay $\kappa/g$ of nonlocal control (left) and local control (right). The blue line and black line represent 90\% fidelity for each case, which just correspond to the single-atom cooperativity
parameter $C\approx290$ and $C\approx350$, respectively.}
\end{figure*}
The significant spontaneous emission of the current system arises from
two upper levels to the other three levels. The rates of
spontaneous emission are $\gamma_1$ from $|e\rangle$ to $|0\rangle$, $\gamma_2$ from
 $|e\rangle$ to $|1\rangle$, $\gamma^{'}_1$ from $|r\rangle$ to $|1\rangle$, and $\gamma^{'}_2$ from
 $|r\rangle$ to $|2\rangle$, with $\gamma_1+\gamma_2=\gamma$ and $\gamma^{'}_1+\gamma^{'}_2=\gamma^{'}$, respectively. These decoherence channels will not vanish, but transform into other forms after the upper levels are adiabatically eliminated (see Appendix A for details). If we suppose $\gamma=\gamma^{'}=2\gamma_1=2\gamma_1^{'}$, the effective spontaneous emission effect in the Lindblad master equation takes the form of
\begin{equation}\label{99}
{\cal L}_{\rm sp}\rho=\sum_{i=1}^3\sum_{j=0}^1{\cal D}[R^i_{je}]\rho+\sum_{i=1}^3\sum_{k=1}^2{\cal D}[R^i_{kr}]\rho,
\end{equation}
where the damping operators
\begin{equation}\label{15}
R_{0e}^i=\sqrt{\frac{\gamma\lambda_a^{i2}}{2\Delta^2}}|0\rangle_{ii}\bigg(\langle0|+\frac{\lambda_b^i}{\lambda_a^i}\langle 1|\bigg),
\end{equation}
\begin{equation}\label{16}
R_{1e}^i=\sqrt{\frac{\gamma\lambda_a^{i2}}{2\Delta^2}}|1\rangle_{ii}\bigg(\langle0|+\frac{\lambda_b^i}{\lambda_a^i}\langle 1|\bigg),
\end{equation}
come from the Raman transition with detuning $-\Delta$, and
\begin{equation}\label{a111}
R_{0e}^i=\sqrt{\frac{\gamma\Omega_a^{i2}}{2\Delta^2}}|0\rangle_{ii}\langle0|,\ \ R_{1e}^i=\sqrt{\frac{\gamma\Omega_a^{i2}}{2\Delta^2}}|1\rangle_{ii}\langle0|,
\end{equation}
and
\begin{equation}\label{a111}
R_{1r}^i=\sqrt{\frac{\gamma\Omega_b^{i2}}{2\Delta^2}}|1\rangle_{ii}\bigg(\langle1|+\frac{\Omega_c^i}{\Omega_b^i}\langle 2|\bigg),
\end{equation}
\begin{equation}\label{a111}
R_{2r}^i=\sqrt{\frac{\gamma\Omega_b^{i2}}{2\Delta^2}}|2\rangle_{ii}\bigg(\langle1|+\frac{\Omega_c^i}{\Omega_b^i}\langle 2|\bigg),
\end{equation}
come from the Raman transition with detuning $\Delta$. Compared with the decoherence factor for a two-level system $\gamma/g$, the rate of effective spontaneous emission in our proposal is reduced to $\gamma\Omega/(g\Delta)$, which makes the current scheme more robust, and provides more feasibility to obtain a high fidelity for different decoherence parameters. In Fig.~\ref{p4}, we plot the contours of fidelities for steady states versus spontaneous emission $\gamma/g$ and cavity decay $\kappa/g$ of nonlocal feedback control (left) and local feedback control (right), and the blue line and black line represent 90\% fidelity for each case, which just correspond to the single-atom cooperativity
parameter $C\approx290$ and $C\approx350$, respectively.

\section{discussion and conclusion}
\begin{figure}
\scalebox{0.22}{\includegraphics{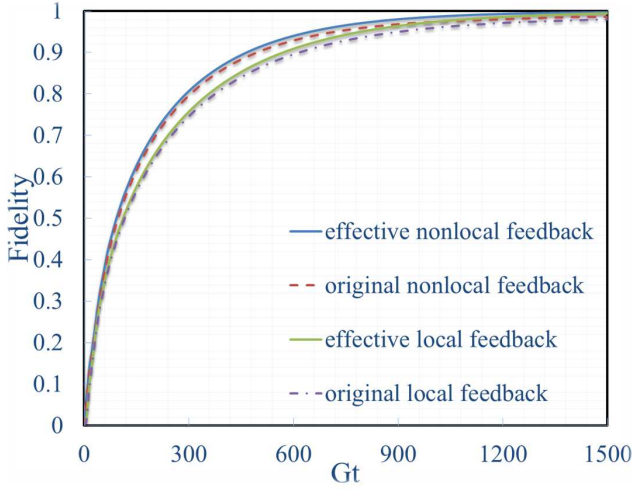} }
\caption{\label{p5}(Color online) The time evolution of fidelities for target state from an initial state $|111\rangle$ with effective master equations compared with the original ones. The corresponding parameters are chosen as $\Omega=0.1G$, $\Delta=200g$, $J=\Delta/\sqrt{2}$, and the driving frequencies are set as $\omega=0.3\pi$ for the nonlocal control, while $\omega=0.5\pi$ for the local control.}
\end{figure}

Actually, the feedback setups are triggered condition on the photons are detected by three detectors, thence the corresponding stochastic master equation reads
\begin{eqnarray}\label{HHHH}
\dot{\rho}&=&-i[{\cal H}_{\rm ac},\rho]+{\cal L }\rho+{\cal L_{\rm sp} }\rho
+\kappa_a{\cal D}[{\bf U_{{\rm fb}}}a]\rho\nonumber\\&&+\kappa_b{\cal D}[{\bf U_{{\rm fb}}}b]\rho+\kappa_c{\cal D}[{\bf U_{{\rm fb}}}c]\rho,
\end{eqnarray}
where
$
{\cal H_{\rm ac}}=G/2|0\rangle_{11}\langle 1|
(c_3+c_2+\sqrt{2}c_1)+G/2|0\rangle_{22}\langle 1|
(c_3-c_2)
+G/2|0\rangle_{33}\langle 1|
(c_3+c_2-\sqrt{2}c_1)+{\rm H.c.}
-\sqrt{2}Jc_1^{\dag}c_1-2\sqrt{2}Jc_2^{\dag}c_2.
$
This master equation can be transformed back to Eqs.~(\ref{H10}) and (\ref{H11}) through an adiabatic
elimination. For a resonator system of Fabry-Perot projected limits, we have the coupling strength between atom and cavity  $g/(2\pi)=770$~MHz, the critical photon number $n_0\equiv\gamma^2/(2g^2)\approx5.7$$\times$$10^{-6}$, and the critical atom number $N_0\equiv2\gamma\kappa/g^2\approx1.9$$\times$$10^{-4}$, which correspond to cavity QED parameters of {\bf(}$g/(2\pi)$,$\kappa/(2\pi)$,$\gamma/(2\pi)${\bf)}=(770~MHz, 21.67~MHz, 2.6~MHz). In order to realize the
effective dissipative dynamics of Eq~(\ref{999}), we set $\Delta=200g$, $J=\Delta/\sqrt{2}$ and $\Omega=0.1G$. By substituting the above parameters into Eq~(\ref{HHHH}), we acquire the time evolution of fidelities for target state from an initial state $|111\rangle$ in Fig.~\ref{p5}, the dashed line and dash-dotted line represent the results from the original nonlocal feedback control and local feedback control, respectively, which are both in conformity with the outcomes obtained from the effective feedback control schemes, indicated by blue line and green line. The fidelities can rise to 98.53\% and 97.98\% at a short time $Gt=1500$, and their values are going to get higher further with the increase of time. One may also pick out another state as the initial state, but will come to an  almost identical conclusion with ours, because our proposal is independent of initial states.

In conclusion, we have presented an efficient scheme for dissipative preparation of tripartite singlet state by trapping atoms into coupled three-cavity arrays. The quantum-jump-based feedback controls are employed to stabilize the quantum system into pure entanglement regardless of initial states. The effects of typical decoherence parameters in cavity QED system are also investigated, which shows that a fidelity outriding 90\% always can be achieved as $C>350$. We believe that our work may open a new avenue for the entanglement preparation experimentally in the near future.
\begin{center}{\bf{ACKNOWLEDGMENT}}
\end{center}

This work is supported by Natural Science Foundation of China under Grant Nos. 11204028, 11405008, and 11547303,
the Fundamental Research Funds for the Central Universities under Grant Nos.  2412016KJ004 and 2412015KJ009, and the Plan for Scientific and Technological Development of Jilin Province under Grant No. 20160520173JH.

\appendix
\section{Derivation of the effective spontaneous emission}
The effective spontaneous emission can be understood with the following toy model. The interaction between classical fields and atom $i$ is described by the Hamiltonian
\begin{eqnarray}\label{a111}
{\cal H}^{i}_{\rm atom}&=&\Omega_a^{i}|e\rangle_{ii}\langle 0|e^{i\Delta t}+
\Omega_b^{i}|r\rangle_{ii}\langle 1|e^{i\Delta t}+\Omega_c^{i}|r\rangle_{ii}\langle 2|e^{i\Delta t}\nonumber\\
&&+\lambda_a^i|e\rangle_{ii}\langle 0|e^{-i\Delta t}+\lambda_b^i|e\rangle_{ii}\langle 1|e^{-i\Delta t}+{\rm H.c.}.
\end{eqnarray}
For a large value of $|\Delta|$, there are two independent transiting channels corresponding detunings $\Delta$ and $-\Delta$, respectively.
Considering only the transition $|0\rangle\leftrightarrow|e\rangle\leftrightarrow|1\rangle$ induced by $\lambda_a$ and $\lambda_b$, with detuning $-\Delta$, the Lindblad master equation, after performing a rotating with respect to $U=\exp(-i\Delta|e\rangle_{ii}\langle e|)$, can be rewritten as
\begin{eqnarray}\label{HHH}
\dot{\rho}^i_{\rm atom}&=&-i[{\cal H}^{i}_{\rm atom},{\rho}^i_{\rm atom}]+\gamma_1{\cal D}\big[|0\rangle_{ii}\langle e|\big]\rho^i_{\rm atom}\nonumber\\
&&+\gamma_2{\cal D}\big[|1\rangle_{ii}\langle e|\big]\rho^i_{\rm atom},
\end{eqnarray}
where the effective Hamiltonian reads
\begin{eqnarray}\label{a111}
{\cal H}^{i}_{\rm atom}=\lambda_a^i|e\rangle_{ii}\langle 0|+\lambda_b^i|e\rangle_{ii}\langle 1|+{\rm H.c.}-\Delta|e\rangle_{ii}\langle e|.
\end{eqnarray}
Explicitly, the field-matrix
elements follows the coupled equations
 \begin{equation}\label{mmm1}
\dot{\rho}^i_{\rm 00}=-i\lambda_a^i({\rho}^i_{\rm e0}-{\rho}^i_{\rm 0e})+\gamma_1{\rho}^i_{\rm ee},
\end{equation}
\begin{eqnarray}\label{mmm2}
\dot{\rho}^i_{\rm ee}&=&-i\lambda_a^i({\rho}^i_{\rm 0e}-{\rho}^i_{\rm e0})
-i\lambda_b^i({\rho}^i_{\rm 1e}-{\rho}^i_{\rm e1})\nonumber\\&&-(\gamma_1+\gamma_2){\rho}^i_{\rm ee},
\end{eqnarray}
\begin{equation}\label{mmm3}
\dot{\rho}^i_{\rm 11}=-i\lambda_b^i({\rho}^i_{\rm e1}-{\rho}^i_{\rm 1e})+\gamma_2{\rho}^i_{\rm ee},
\end{equation}
\begin{eqnarray}\label{mmm4}
\dot{\rho}^i_{\rm 0e}&=&-i\lambda_a^i({\rho}^i_{\rm ee}-{\rho}^i_{\rm 00})
+i(\Delta{\rho}^i_{\rm 0e}+\lambda_b^i{\rho}^i_{\rm 01})\nonumber\\&&-\frac{\gamma_1+\gamma_2}{2}{\rho}^i_{\rm 0e},
\end{eqnarray}
\begin{eqnarray}\label{mmm5}
\dot{\rho}^i_{\rm 1e}&=&-i\lambda_b^i({\rho}^i_{\rm ee}-{\rho}^i_{\rm 11})
+i(\Delta{\rho}^i_{\rm 1e}+\lambda_a^i{\rho}^i_{\rm 10})\nonumber\\&&-\frac{\gamma_1+\gamma_2}{2}{\rho}^i_{\rm 1e},
\end{eqnarray}
\begin{equation}\label{mmm6}
\dot{\rho}^i_{\rm 01}=-i(\lambda_a^i{\rho}^i_{\rm e1}-\lambda_b^i{\rho}^i_{\rm 0e}).
\end{equation}
Now we make the assumption that both $\dot{\rho}^i_{\rm 0e}=0$ and $\dot{\rho}^i_{\rm 1e}=0$, then the values of
${\rho}^i_{\rm 0e}$ and ${\rho}^i_{\rm 1e}$ are found to be
\begin{equation}\label{HHH}
{\rho}^i_{\rm 0e}=-\frac{2i[\lambda_a^i({\rho}^i_{\rm ee}-{\rho}^i_{\rm 00})-\lambda_b^i{\rho}^i_{\rm 01}]}{\gamma_1+\gamma_2-2i\Delta},
\end{equation}
\begin{equation}\label{HHH}
{\rho}^i_{\rm 1e}=-\frac{2i[\lambda_b^i({\rho}^i_{\rm ee}-{\rho}^i_{\rm 11})-\lambda_a^i{\rho}^i_{\rm 10}]}{\gamma_1+\gamma_2-2i\Delta}.
\end{equation}
Substituting these results
into  Eqs.~(\ref{mmm1}), (\ref{mmm2}), (\ref{mmm3}) and (\ref{mmm6}), and adiabatically eliminating the excited state $\rho_{ee}$, we have
\begin{eqnarray}\label{HHH}
\dot{\rho}^i_{\rm 00}&=&-i\frac{\lambda_a^i\lambda_b^i}{\Delta}({\rho}^i_{\rm 01}-{\rho}^i_{\rm 10})-\frac{\gamma_2^2}{\Delta^2}\lambda_a^{i2}\rho_{00}^i+\frac{\gamma_1^2}{\Delta^2}\lambda_b^{i2}\rho_{11}^i\nonumber\\
&&+\frac{\lambda_a^i\lambda_b^i}{2\Delta^2}(\gamma_1-\gamma_2)(\rho^i_{01}+\rho^i_{10}),
\end{eqnarray}
\begin{eqnarray}\label{HHH}
\dot{\rho}^i_{\rm 01}&=&-i\frac{1}{\Delta}[\lambda_2^i(\lambda_1^i\rho^i_{00}+\lambda_2^i\rho^i_{01})
-\lambda_1^i(\lambda_1^i\rho^i_{01}+\lambda_2^i\rho^i_{11})]\nonumber\\
&&-\frac{\gamma_1+\gamma_2}{2\Delta^2}[\lambda_1^i\lambda_2^i(\rho^i_{00}+\rho^i_{11})\nonumber\\&&
+(\lambda_a^{i2}+\lambda_b^{i2})\rho^i_{01})],
\end{eqnarray}
\begin{eqnarray}\label{HHH}
\dot{\rho}^i_{\rm 11}&=&-i\frac{\lambda_a^i\lambda_b^i}{\Delta}({\rho}^i_{\rm 10}-{\rho}^i_{\rm 01})-\frac{\gamma_1^2}{\Delta^2}\lambda_b^{i2}\rho_{11}^i+\frac{\gamma_2^2}{\Delta^2}\lambda_a^{i2}\rho_{00}^i\nonumber\\
&&+\frac{\lambda_a^i\lambda_b^i}{2\Delta^2}(\gamma_2-\gamma_1)(\rho^i_{01}+\rho^i_{10}),
\end{eqnarray}
which just recover the result of Eqs.~(\ref{15}) and (\ref{16}) as $\gamma_1=\gamma_2$.


\begin{thebibliography}{999}
\bibitem{ein} A. Einstein, B. Podolsky, and N. Rosen, Phys. Rev. {\bf47},
777 (1935).
\bibitem{Bell} J. S. Bell, Physics (Long Island City, N.Y.) \textbf{1}, 195 (1965).
\bibitem{woo}W. K. Wootters
Phys. Rev. Lett. {\bf80}, 2245 (1998).
\bibitem{vidal} G. Vidal and R. F. Werner, Phys. Rev. A {\bf65}, 032314 (2002).
\bibitem{ghz} D. M. Greenberger, M. A. Horne, and A. Zeilinger, 1989,
{\it Going beyond Bell¡¯s theorem in Bell¡¯s Theorem, Quantum
Theory, and Conceptions of the Universe} (Kluwer Academic,
Dorthecht)
\bibitem{wstate}W. D\"{u}r, G. Vidal, and J. I. Cirac,  Phys. Rev. A {\bf62}, 062314 (2000).
\bibitem{HJ}H. J. Briegel and R. Raussendorf, Phys. Rev. Lett. {\bf86},
910 (2001); R.
Raussendorf and H. J. Briegel, Phys. Rev. Lett. {\bf86}
5188 (2001).
\bibitem{Cabe}A. Cabello, Phys. Rev. Lett. {\bf89}, 100402 (2002).
\bibitem{Cabe1}A. Cabello, J. Mod. Opt. {\bf50}, 10049 (2003).
\bibitem{Jin}G. S. Jin, S. S. Li, S. L. Feng, and H. Z. Zheng, Phys.
Rev. A {\bf71}, 034307 (2005).
\bibitem{Lin}G. W. Lin, M. Y. Ye, L. B. Chen, Q. H. Du, and X. M. Lin, Phys. Rev. A {\bf76}, 014308 (2007).
\bibitem{shao}X. Q. Shao, H. F. Wang, L. Chen,
S. Zhang, Y. F. Zhao, and K. H. Yeon, New J. phys. {\bf12}, 023040 (2010).
\bibitem{chenz} Z. Chen,Y. H. Chen, Y. Xia, J. Song, and B. H. Huang, Sci. Rep. {\bf6}, 22202 (2016).




\bibitem{ple} M. B. Plenio, S. F. Huelga, A. Beige, and P. L. Knight
Phys. Rev. A {\bf59}, 2468 (1999); A. Beige, S. Bose, D. Braun, S. F. Huelga, P. L. Knight, M. B. Plenio, and V. Vedral, J. Mod. Opt. {\bf47}, 2583(2000); P. Horodecki,
Phys. Rev. A {\bf63},022108(2001).
\bibitem{ple1} M. B. Plenio and S. F. Huelga
Phys. Rev. Lett. {\bf88}, 197901 (2002).
\bibitem{yi} X. X. Yi, C. S. Yu, L. Zhou, and H. S. Song
Phys. Rev. A {\bf68}, 052304 (2003).
\bibitem{kastoryano}M. J. Kastoryano, F. Reiter, and A. S. S{\o}rensen, Phys. Rev. Lett. {\bf106}, 090502 (2011).
\bibitem{busch}J. Busch, S. De, S. S. Ivanov, B. T. Torosov, T. P. Spiller, and A. Beige, Phys. Rev. A {\bf84}, 022316 (2011).
\bibitem{shen}L. T. Shen, X. Y. Chen, Z. B. Yang, H. Z. Wu, and S. B. Zheng, Phys. Rev. A {\bf84}, 064302 (2011).

\bibitem{torre}E. G. Dalla Torre, J. Otterbach, E. Demler, V. Vuletic, and M. D. Lukin, Phys. Rev. Lett. {\bf110}, 120402 (2013).
\bibitem{rao}D. D. Bhaktavatsala Rao and K. M{\o}lmer, Phys. Rev. Lett. {\bf111}, 033606 (2013).
\bibitem{carr} A. W. Carr and M. Saffman, Phys. Rev. Lett. {\bf111}, 033607 (2013).
\bibitem{lin} Y. Lin, J. P. Gaebler, F. Reiter, T. R. Tan, R. Bowler, A. S. S{\o}ensen, D. Leibfried, and D. J. Wineland, Nature (London) {\bf 504}, 415 (2013).
\bibitem{zz}Z. Leghtas, U. Vool, S. Shankar, M. Hatridge, S. M. Girvin,
M. H. Devoret, and M. Mirrahimi, Phys. Rev. A {\bf88}, 023849
(2013).
\bibitem{shaoxq} X. Q. Shao, T. Y. Zheng, and S. Zhang, Phys. Rev. A
{\bf85}, 042308 (2012); X. Q.  Shao, T. Y. Zheng, C. H. Oh, and S. Zhang,
Phys. Rev. A {\bf89}, 012319 (2014); X. Q.  Shao, J. B. You, T. Y. Zheng, C. H. Oh, and S. Zhang,
Phys. Rev. A {\bf89}, 052313 (2014).
\bibitem{ben}C. D. B. Bentley, A. R. R. Carvalho, D. Kielpinski, J. J. Hope, Phys. Rev. Lett. {\bf113}, 040501 (2014).
\bibitem{gon}A. Gonz\'{a}lez-Tudela, V. Paulisch, D. E. Chang, H. J. Kimble, and J. I. Cirac, Phys. Rev. Lett. {\bf115}, 163603 (2015).



\bibitem{1}A. R. R. Carvalho and J. J. Hope, Phys. Rev. A {\bf76}, 010301(R)
(2007).
\bibitem{2}A. R. R. Carvalho, A. J. S. Reid, and J. J. Hope, Phys. Rev. A
{\bf78}, 012334 (2008).
\bibitem{3}R. N. Stevenson, J. J. Hope, and A. R. R. Carvalho,  Phys. Rev. A
{\bf84}, 022332 (2011).

\bibitem{wang}J. Wang, H. M. Wiseman, and G. J. Milburn, Phys. Rev. A {\bf71},
042309 (2005).



\bibitem{Nielsen}M. A. Nielsen and I. L. Chuang, \emph{Quantum Computation and Quantum Information} (Cambridge
University Press, Cambridge, 2000).

\bibitem{spi}  S. M. Spillane, T. J. Kippenberg, K. J. Vahala,  K. W. Goh, E. Wilcut,
and H. J. Kimble, Phys. Rev. A {\bf71}, 013817 (2005).

\end{thebibliography}
\end{document}